# Ranges of Atmospheric Mass and Composition of Super Earth Exoplanets

Linda T. Elkins-Tanton and Sara Seager

Dept. Earth, Atmospheric, and Planetary Sciences
MIT
77 Massachusetts Ave
Cambridge MA 02139

Linda T. Elkins-Tanton: Building 54-824, ltelkins@mit.edu
Sara Seager (co-appointed in Dept. of Physics): Building 54-1626, seager@mit.edu

## Abstract

Terrestrial-like exoplanets may obtain atmospheres from three primary sources: Capture of nebular gases, degassing during accretion, and degassing from subsequent tectonic activity. Here we model degassing during accretion to estimate the range of atmospheric mass and composition on exoplanets ranging from 1 to 30 Earth masses. We use bulk compositions drawn from primitive and differentiated meteorite compositions. Degassing alone can create a wide range of masses of planetary atmospheres, ranging from less than a percent of the planet's total mass up to ~6 mass% of hydrogen, ~ 20 mass% of water, and/or ~5 mass% of carbon compounds. Hydrogen-rich atmospheres can be outgassed as a result of oxidizing metallic iron with water, and excess water and carbon can produce atmospheres through simple degassing. As a byproduct of our atmospheric outgassing models we find that modest initial water contents (10 mass% of the planet and above) create planets with deep surface liquid water oceans soon after accretion is complete.

## Introduction

Planets have three main opportunities for obtaining an atmosphere: capture from the nebula, degassing during accretion, and later degassing from tectonic processes. While capture of gases is likely to be a critical process with gas giant planets, its importance for terrestrial planets is unproven. Low-mass terrestrial planets are unable to capture and retain nebula gases, and nebular gases may have largely dissipated from the inner solar system by the time of final planetary accretion.

Here we focus on the range of atmospheric masses possible from degassing terrestrial-analog materials in the planetary accretion process. We further calculate the mass of volatiles retained in the planetary interior and available for later degassing through tectonic processes. Atmospheric mass and composition for terrestrial planets is therefore closely related to the composition of the solid planet. These simple end-member models will produce a framework for further analysis.

The library of meteorites that have fallen to Earth provide a range of plausible starting compositions for planetary accretion and degassing. The chondrites are the least processed and therefore most primitive meteorite class. Chondrites contain a variety of silicate components, including chondrules (perhaps the earliest condensed material from the planetary nebula), minerals such as olivine and pyroxene, silicate glass, and in some cases water- or carbon-rich veins, indicating alteration after formation. In these meteorites, water is most often in the form of OH within a silicate mineral crystal. The reason the OH within the mineral is called water is that it likely existed as water when the mineral formed and if the mineral melts the OH is released as water. Many chondrites have low oxygen contents and thus also contain metallic iron and nickel. The chondrites with the most water contain little or no metallic iron, while chondrites with little or no water can contain metallic iron and nickel approaching 50 percent by mass (50 mass%) (Hutchison 2004). Wood (2005) reports up to 20 mass% of water in carbonaceous chondrites.

A second class of meteorites pertinent to planet atmosphere formation is the achondrites. Achondrites generally lack metallic iron and are thought to represent the silicate remnants of planetesimals that accreted and differentiated into metallic iron cores and silicate mantles, only to be broken apart in later collisions. Achondrites have far lower water contents than do most chondrites. Jarosewich (1990) reports that achondrites can contain up to ~3 mass% water, though many have no water at all.

The terrestrial planets were originally thought to have accreted from primordial undifferentiated disk material analogous to chondritic meteorites (*e.g.*, Ringwood 1979). Ringwood suggested that the terrestrial planets were built of a reduced, volatile-free components and a second, volatile-rich component, to explain the reduced metallic core and the oxidized mantle and surface. Wänke & Dreibus (1988) also espoused a two-component accretion process, but suggested that the reduced component was an early component, and that as accretion proceeded, more and more of the oxidized component was added. They further suggest that the reduced component formed closer to the Sun than Mars, and that the oxidized component originated at greater radii.

More modern simulations of accretion indicate that planets form from differentiated planetary embryos on the scale of thousands of kilometers in radius, and that these embryos move radially in the solar system during accretion and so form planets that are mixtures of material from the inner and outer disk (*e.g.*, Wiedenschilling 1977; Kokubo & Ida 2000; Raymond, Quinn, & Lunine 2006). Even assuming planetary formation from differentiated material only, the range of possible bulk compositions for terrestrial planets in our solar system is extremely wide.

Two extreme end members of possible bulk compositions are rocky, iron metal-bearing meteorites with no water, and mixtures of metallic iron, silicate rock, and unconstrained volumes of outer solar system volatiles and ices. We seek to explore the wide variety of planetary atmospheres that can be created from oxidized or reduced accreting material, as well as the processes that might lead to their initial production. This is in contrast to explaining the differential compositions of the terrestrial planets' metallic cores and oxidized silicate mantles, as Wänke & Dreibus (1988) have done.

In Earth, Venus, and Mars, present-day atmospheres indicate that water and carbon dioxide have remained since accretion, though perhaps the bulk of volatiles were delivered toward the end of accretion (*e.g.*, Morbidelli et al. 2000; Chambers 2006; Raymond *et al*. 2006; O'Brien, Morbidelli, & Levison 2006). In the solar system terrestrial planets, therefore, not all of the water reacted with iron; these planets have both a metallic core and a volatile-rich atmosphere. For later stages of accretion, after core formation, reactions may be limited to the mantle and atmosphere and not involve the core.

In this paper we calculate possible atmospheric masses and compositions resulting from planets accreted from planetesimals of different meteoritic compositions. In section 2 we describe our models, in section 3 we present our results, in section 4 we discuss the implication of the results and in section 5 we conclude with a summary.

## Models

Four models arise from the accretion scenarios described in the Introduction (see Table 1 and Figure 1). In the first case we consider planets accreted from chondrites in which water and metallic iron are allowed to react, and the product hydrogen outgasses. In the second case we consider planets after core formation by differentiation in which achondritic compositions undergo reactions that only involve rock in the mantle and gas in the atmosphere and do not involve the metallic core.

In our models we primarily consider the interactions of water, metallic iron and silicates. The bulk of a terrestrial planet is iron and silicate. Water is the most abundant atmosphere-forming component that also interacts with the planetary interior. In addition to water, carbon is the second most abundant atmosphere-forming component in meteorites, but it is much less reactive with silicates and iron.

Planets in our models may be built from:

- **1A. Primitive material alone.**

  Primitive (chondritic) material accretes into a planet. Water oxidizes metallic iron until available water is exhausted. The by-product of the water-iron reaction is degassed hydrogen. In the rare case that all of the metal becomes oxidized before water is depleted, water will also be degassed. In this case we take actual chondrite compositions with their existing water and metallic iron content.
- **1B. Primitive material with added water.**

Primitive material accretes into a planet as in case 1A, but with additional water exactly sufficient to oxidize all iron. This situation is an end-member model in which all metallic iron is oxidized.

- **2A. Differentiated material alone**.

  Achondritic material accretes to a protoplanet with an existing core. In this scenario the silicate mantle of the planet is assumed to be fully melted (a whole-mantle magma ocean) and then solidifies, partitioning water between the solidifying mantle minerals, the evolving liquids, and the growing atmosphere.

- **2B. Differentiated material with added water.**

  Achondritic material accretes with additional volatiles. This case is similar to case 2A but with additional volatiles added during the magma ocean phase.

For the planets built of primitive (chondritic) material in models 1A and 1B, we simply calculated algebraically the amount of water available to react with metallic iron, and the resulting degassed hydrogen and water atmosphere. The only chemical reactions in these models are between water and metallic iron; no melting, recrystallization, or incorporation of water into silicate minerals is included. The exclusion of these more complex processes allows the simplicity of the calculation used here, and makes these models end-members. Average compositions of the 13 classes of chondrites from Hutchison (2004) are used in calculations of water and iron reaction. Compositions are divided into silicate, metallic iron, and water fraction.

The planets built from differentiated (achondritic) material (cases 2A and 2B) require more careful and complex calculation than the models we used for planets built out of chondritic material (cases 1A and 1B) because the energy of accretion of large differentiated bodies implies some fraction of melting of the protoplanet. If the planet has melted, its process of solidification will determine the composition and mass of the outgassed atmosphere.

We assume the bulk composition of planets built from differentiated (achondritic) material (cases 2A and 2B) are similar to the Earth, Venus, and Mars, that is, that they consist of an iron core of about half the radius of the solid planet, surrounded by a silicate mantle, and contain a volatile load that consists primarily of water. We use the achondritic water contents from Jarosewich (1990). No initial volatile loss to space is assumed: all the volatiles in the initial bulk composition remain with the planet, either incorporated into the solid material or degassed into the atmosphere. Similarly, we assume no water loss to space as the atmosphere grows over the brief time required to solidify a planetary magma ocean (from a hundred thousand to a few millions of years, Abe 1997; Elkins-Tanton, 2008). Planets with 1, 5, 10, 20, and 30 times the mass of the Earth are modeled with initial water contents of 0.5, 1, 5, and 10 mass% water in the bulk silicate magma ocean. Water contents up to 3 mass% have been found in achondrites; above this value the water is considered a later addition (these latter are models 2B, differentiated material with added water).

Processes of magma ocean solidification and the details of these models are presented and discussed in Elkins-Tanton, Parmentier, & Hess (2003), and Elkins-Tanton & Parmentier (submitted). We assume the magma ocean solidifies from the bottom upward, because the steep slope of the liquid adiabat first intersects the shallower slope of the solidus at depth. Volatiles are enriched in evolving magma ocean liquids as solid cumulates form and liquid volume decreases (though a small fraction of volatiles are incorporated into the solids, as discussed below). Volatiles in excess of the saturation capacity of the magma will degas into the atmosphere. As a limiting case, we assume that volatiles will partition between the atmosphere and the magma ocean liquids according to their equilibrium partial pressures. We fitted the following equations for the atmospheric partial pressures of water $p_{H2O}$and carbon dioxide $p_{CO2}$ [Pa] as a function of the magma ocean dissolved water $H_2O_{magma}$ and carbon dioxide $CO_{2magma}$ [mass% in the liquid] contents from data from Papale (1997):

$$p_{H_2O} = \left[\frac{H_2O_{magma} - 0.30}{2.08 \times 10^{-4}}\right]^{\frac{1}{0.52}} \text{ [Pa], and} \qquad (1)$$

$$p_{CO_2} = \left[\frac{CO_{2\,magma} - 0.05}{2.08 \times 10^{-4}}\right]^{\frac{1}{0.45}} \text{ [Pa].}$$

Water is retained in the solidifying planet as well as degassed into the atmosphere. While the atmosphere grows from degassing and the magma ocean liquids evolve as degassing and solidification proceed, the nominally anhydrous silicate mantle minerals can incorporate a dynamically and petrologically significant amount of OH (as much as 1,000-1,500 ppm for olivine and orthopyroxene (*e.g.*, Koga et al. 2003; Bell et al. 2004; Hauri, Gaetani, & Green 2006).

We calculate mineral compositions in equilibrium with the magma ocean composition using experimental distribution coefficients and saturation limits. Rates of equilibrium crystal growth are generally far faster than rates of solidification, so the assumption of equilibrium is reasonable. At pressures higher than 90 GPa the solidifying mineral is post-perovskite phase. From 90 to 22 GPa, perovskite and magnesiowustite are assumed to crystallize. From 22 to 15, γ-olivine (ringwoodite) and majorite are stable, and from 15 GPa to the planetary surface the phase assemblage is garnet, clinopyroxene, orthopyroxene, and olivine. Since each of these minerals accepts OH in a different ratio with the coexisting silicate liquid, and is saturated at a different limiting value, the internal pressure range of the planet strongly determines the mass of water retained rather than degassed. For further details of water partitioning see Elkins-Tanton & Parmentier (2005) and Elkins-Tanton & Parmentier (submitted).

## Results

### *Primitive (chondritic) starting compositions: Model 1A*

A range of water, water + hydrogen, and hydrogen atmospheres are produced in these models. Assuming a planet accretes from only one meteorite class, we can explore extreme end members (see Figures 2 and 3). Built from the meteorite class CI, a planet's maximum water mass percentage could be as high as 23%. Built from the meteorite class EH a planet would have an atmosphere consisting of 0.4 mass% of the planet hydrogen and 0.1 mass% carbon; this class results in an atmosphere with the highest degassed hydrogen content in a planet that degasses no water. Planets forming from other chondrite classes would have atmospheres that consist of water, hydrogen and carbon. One such class, the CRs, would degas as much as 0.9 mass% of the planet hydrogen (Table 2). Planetary accretion modeling efforts suggest that planets are formed from a mixture of planetesimal types, and so the pure end members are illustrative only.

In Model 1A, the initial iron fraction controls the atmospheric composition. Planets formed from the meteorite classes listed in Figure 2 will not all have iron cores. In some cases there will not be enough metallic iron that escapes oxidation, and therefore not enough metallic iron to form a core. In the cases where insufficient metallic iron remains to make a core, the atmosphere will consist of water ± hydrogen. Conversely, because the primitive compositions with the most metallic iron also possess the least water, the production of a hydrogen atmosphere through iron oxidation is limited for high-metal compositions (Figures 2, 3). If sufficient metallic iron remains after reaction to make a core the size of those of the terrestrial planets, then in each case water has been completed dissociated through oxidation and the planetary atmosphere consists of hydrogen without water.

*Primitive (chondritic) starting compositions with added water: Model 1B*

With the addition of sufficient water to oxidize all the metallic iron in the starting composition, hydrogen atmospheres of as much as 6 mass% of the planet can be produced. This maximum hydrogen atmosphere is produced by oxidation of the subclass of chondrites known as EH, or high-iron chondrites.

In addition to water and hydrogen, some of the chondrite classes contain significant amounts of carbon. Carbon content of atmospheres of between 0.1 and 5 mass% of the planetary mass are possible, based on the assumption that all of the carbon is volatilized (Figure 3).

*Differentiated (achondritic) starting compositions: Model 2A*

In Models 2A and 2B, achondritic starting compositions are processed through a magma ocean and volatiles are partitioned between the silicate mantle and the atmosphere. Because the magma ocean process depends on pressure ranges and mineral content, we must discuss results in terms of mass-specific planet models (Figure 4). Our calculations are consistent with observations of Earth's volatile budget, specifically their partition between the mantle and the atmosphere.

If the terrestrial-like exoplanets are built from materials similar to the terrestrial planets in our solar system (i.e., the achondrites without added water), our results provide an estimate for how much water might be present in the atmosphere. As an example a 10-Earth mass planet with only achondritic water contents (<3 mass %) can outgas ~ 0.2 Earth masses of water into its atmosphere (Figure 4). With no added volatile-rich accreting material during planet formation, 3% is the maximum available from achondritic starting compositions, but less than half this water content is more likely. Although the water atmospheres produced in these models are smaller than some of the water atmospheres produced in model 1A, these are more than sufficient to explain, for example, Earth's water budget.

Carbon is far less able to partition into mantle materials than is water. To a close first approximation, all carbon will rapidly degas into the growing planetary atmosphere. Jarosewich (1990) reports achondritic carbon contents from 0 to ~3 mass%, though the majority of achondritic compositions have on the order of 0.01 mass%. A far higher fraction of carbon will be degassed than will water (Figure 5).

When processed through a solidifying magma ocean, between ~70 and ~97% of volatiles are degassed, while the remainder are held in mantle silicate minerals (these models do not include any retained interstitial melt). The larger the planet and the lower the initial water fraction, the higher the water fraction retained in the planet. Though the Earth likely began with not more than 1 mass % of water, the mass of volatiles remaining in the silicate mantle after a hypothetical magma ocean is still geodynamically significant at tens to hundreds of parts per million (Elkins-Tanton & Parmentier, submitted). Silicate mantle convection and magmatism is greatly facilitated by both a lowering of solid-state viscosity and a lowering of the melting temperature of the silicates even at these volatile contents, and with later magmatism, ~99% of the water in the melting source region will be released into the atmosphere.

When solidified from a magma ocean the highest water content the silicate minerals retain in these models is about 3 mass% of the minerals; this occurs toward the end of solidification when water is maximally enriched in the remaining liquids, and when the mineral assemblage is uniquely able to incorporate more water (the remainder of the newly solidified silicate mantle contains far less water). We emphasize that the high 3 mass% value is not the uniform value for the whole mantle, but only the highest value found among the great variation caused by varying liquid water contents and varying crystallizing assemblages throughout magma ocean solidification. This 3 mass% value is consistent with the 3 mass% maximum water content of the achondrite meteorites themselves: since achondrites likely represent the silicate mantles of differentiated planetary embryos, they may themselves be the results of earlier magma ocean processing.

*Differentiated (achondritic) starting compositions with added water: Model 2B*

We now turn to Model 2B, which like 2A involves achondritic material processed through a magma ocean, but with additional accretion of volatile material. Models of planetary accretion indicate that volatile-rich material would migrate inward from the outer solar system but quantities are unconstrained. We therefore choose magma oceans of initial water content of 5% and 10%., which represents additions of 2% and 7% over the maximum water present in achondritic material. The results for Model 2B are shown alongside the results for Model 2A in Figure 4.

As an example, a 10 Earth-mass planet with 5 mass% initial water in its silicate fraction degasses ~ 0.3 Earth masses of water. We emphasize that with accretion of a modest amount of volatile-rich material, massive water atmospheres can be created on rocky planets. If the volatile-rich accreting material was half water by mass and the remainder of the silicate magma ocean was accreted from achondritic material with 3 mass% water, only 4% of the mantle would have to be composed of the high-volatile incoming material to produce a magma ocean with 5 mass% water.

A very significant finding of this work is that above a certain initial water content of the magma ocean, the planet will immediately form a fluid ocean of either super critical water or liquid water, rather than an entirely gaseous water atmosphere. This happens because the magma ocean becomes so enriched in water that silicate solidification ends and the water remains on the planetary surface. Silicate solidification ends because the few oxides remaining are soluble in water and are stable as minor constituents in the water ocean. Recall from Section 2 that magma oceans solidify from the bottom upward and so this final water-rich fluid develops at the surface.

As an example, a 10 Earth-mass planet that begins with more than ~18 mass% of water in its silicate fraction, and a high heat of accretion, is left at the end of solidification with a water or critical fluid ocean under a very high partial pressure of water in its atmosphere. A 20 Earth-mass planet forms a surface water layer at an initial silicate fraction water content above ~12 mass%, and a 30 Earth-mass planet above ~10 mass%. This is one route to an ocean planet.

## Discussion

### *Atmospheric Composition*

Our models result in degassing of hydrogen, water, and carbon compounds that include oxygen. Once these species are in the planet atmosphere, several interrelated atmospheric processes will determine the final chemical composition of the planet. These processes include atmospheric escape, photolysis of molecules, and chemical kinetics (rates of chemical processes).

Atmospheric escape (thermal and/or nonthermal) is the most significant process in shaping an atmosphere from its initial mass and composition to its final form. For example, hydrodynamic escape of hydrogen would also drag off heavier molecules. Planets more massive than the Earth with higher surface gravities and planets receiving less stellar irradiation than Earth may be able to retain hydrogen and avoid atmospheric escape. In Figure 6, we show a comparison of a planet's escape velocity compared to the thermal velocity of different elements at the exospheric temperature. While representing only thermal escape, this figure serves to illustrate the possibility that massive super Earths can retain hydrogen against thermal escape.

To estimate the theoretical atmospheric composition of a specific exoplanet, in the absence of observations, one would have to consider detailed models of photochemistry and atmospheric escape, coupled to a lower atmosphere model of the planet's temperature-pressure distribution with altitude. These models would depend on the planet's surface gravity and exospheric temperature, the latter of which depends upon the planet-star orbital separation, the star type, and the upper atmosphere models. In this work we aim for a first-order description of the range of masses and compositions of exoplanet super Earth atmospheres from degassing. Instead of focusing on individual exoplanets with a given mass, orbit and host star type, we explore three generic end-member scenarios.

The first case we consider is that where all hydrogen has escaped. Photolysis would destroy hydrogen-bearing molecules and the subsequent escape of hydrogen to space would prevent them from reforming. In the absence of a liquid water ocean the atmosphere would be dominated by $CO_2$, (like Venus's atmosphere). Planets with a liquid water ocean would also have atmospheric $H_2O$, and some of the $CO_2$ may be sequestered in the ocean.

In the second case we consider a planet where hydrogen is retained in the planet atmosphere, but the atmosphere conditions (low temperatures and pressures) are such that the reaction rates to form $CH_4$ and $NH_3$ are very slow. This chemical kinetic "bottleneck" is important because in a hydrogen-rich environment in thermodynamic equilibrium $CH_4$ is the dominant form of carbon and $NH_3$ is the dominant form of nitrogen. In this hydrogen-rich atmosphere limited by chemical kinetics, the $CH_4$ or $NH_3$ would be destroyed by photolysis and prevented from reforming. In such a planet atmosphere, the dominant molecules of H, C, O, N, would be $H_2$, CO, $H_2O$, and $N_2$. The mixing ratios will depend upon the abundances of each element. This description is in agreement with Hashimoto, Abe, & Sugita (2007), who find that the early atmospheres of the Earth and Venus are likely to be reducing atmospheres dominated by carbon species.

The third end-member case we consider is one where again hydrogen does not escape to space, owing to the planet's mass and amount of received irradiation. In this third case we consider atmospheric conditions where the temperature and/or pressure are high enough to permit the formation of $CH_4$ and $NH_3$,. The solar system giant planets all have abundant $CH_4$. Even though photolysis of $CH_4$ to form higher hydrocarbons is a significant process on the solar system giant planets, the high temperatures and pressures deep in the giant planets' atmospheres enable $CH_4$ to reform. While in this work we are not considering giant planet atmospheres, we have shown that atmospheric masses could be well above 1 to 10 percent of the planet's mass. Such massive atmospheres are likely to be hot at depth, and because the atmosphere turnover time is short there will be some planets where chemical kinetics are not a limiting factor, enabling the stable existence of $CH_4$ and $NH_3$.

This third case therefore represents one with massive planets and massive atmospheres. The dominant form of H, C, O, N would be $H_2$, $H_2O$, $CH_4$ and $NH_3$. At high temperatures (over several hundred degrees) CO may also be present and the ratio of $CO/CH_4$ will depend on the temperature distribution in the atmosphere and on the C/O ratio.

The most massive atmospheres predicted by our models are dominated by water (Figures 2 – 4). Some chondritic initial compositions would create planets with a water atmosphere upward of 20 mass% of the planet. Though carbon can form a significant portion of the atmosphere, even to the point of carbon species making up the entire mass of the atmosphere, carbon in the atmosphere will not account for more than 5 mass% of the planet. Interestingly, these models indicate that very little nitrogen would be delivered to a young terrestrial planet by chondritic meteorite compositions. The maximum ~0.2 mass% of nitrogen from chondrites, however, is sufficient to create the Earth's nitrogen atmosphere as long as water condenses into oceans and carbon is sequestered by geologic processes.

The atmospheres discussed here result from planetary accretion from a specific meteoritic analog. This end-member outcome is highly unlikely, however; planets in this and other solar systems likely accrete from material originating at a range of radii in the planetary disk (*e.g.*, Wiedenschilling 1977; Kokubo & Ida 2000; Raymond et al. 2006). Additionally, the library of meteorites available on Earth is unlikely to cover the entire range of planet-building material from the early solar system, and may not be characteristic of other metal-rich planetary disks. Attempts to fit the solid compositions of the Earth and Mars to those of meteorites demonstrates that the planets are made of mixtures of compositions, and that no one meteorite composition is sufficient (Burbine & O'Brien 2004). Degassed atmospheres in young rocky planets are therefore likely to be mixtures of or even outside the range of compositions considered here.

*The special problem of helium*

While reasonably massive hydrogen atmospheres can be formed through accretionary degassing, helium appears to be more evasive. Helium is not incorporated into silicate minerals in any significant amount; it partitions into crystallizing silicate minerals at one ppm or less (Heber et al. 2007). Helium may be reasonably assumed to degas almost in its entirety during energetic accretion. Harper & Jacobsen (1996) summarize measurements of helium in chondritic meteorites. The data indicate that even in the most optimistic compositional scenarios primitive material as measured in chondritic meteorites can make an atmosphere of only ~ $10^{-11}$ mass % of the planet. Though the meteorites may have significantly degassed between their formation and their eventual fall to Earth and measurement, these results nonetheless imply that large helium atmospheres must be formed through mechanisms other than degassing, or from starting materials other than those we considered. One of our main findings is that atmospheres created from outgassing will lack helium in any significant quantity. As on Earth, helium will come almost entirely from radioactive decay of heavier elements. If there is a way to observe He in hydrogen-rich exoplanet atmospheres, one could distinguish between atmospheres created from outgassing alone and atmospheres formed from accretion of nebular gases.

*Arguments in favor of iron oxidation*

Hydrogen atmospheres in these models result from outgassing following oxidation of metallic iron by reaction with water. The existence of metallic cores in terrestrial planets in our solar system implies that the earliest accretion occurred under reducing conditions, because not all of the metallic iron was oxidized. In other words, the earliest, innermost planetary nebula may have been significantly lacking in oxygen. The absence of oxygen is critical to preserving iron in the metallic state; in contrast the presence of oxygen creates a mixture of iron oxides and iron metal. Water may be the most important oxygen-bearing agent.

Debate exists in the planetary community about the oxidation of iron in magma oceans, i.e., the fate of water and other volatiles in a hot accreting planet. The water might be in the form of liquid water in the planet or OH in silicate magma. The reaction between water and metallic iron liberates hydrogen, allowing hydrogen alone to escape into the atmosphere (Wänke & Dreibus 1994). In small planets, hydrogen would be expected to leave the atmosphere quickly through hydrodynamic escape (e.g., Hunten, Pepin, & Walker 1987). Some authors favor this case in which no water will remain in the mantle or atmosphere after planetary formation.

The presence of water on Earth (and other terrestrial planets) as well as some chondritic and achondritic meteorites indicates that at least some accretionary processes do not destroy water. In fact we see evidence for a range of reducing conditions in the primitive (chondritic) meteorites, pointing to different amounts of intial water. Those chondrites with sufficient metallic iron to form a planetary core lack water, and those with significant water content have little to no metallic iron. We therefore assert that planets form with a range of final water contents. Therefore, a range of metallic core fractions and outgassed H fractions are a natural consequence.

Our hydrogen-outgassed atmosphere relies on metallic iron being oxidized by reaction with water. In a molten planet there is a competition between metallic iron sinking to the core and oxidation reactions with water. If water and metallic iron come into contact at the high temperatures of a magma ocean, iron oxidation occurs immediately, and its rate is controlled by diffusion of hydrogen away from the metal, and diffusion of oxygen into the metal. At magmatic temperatures iron metal may be oxidized to a depth on the order of 0.5 cm over a time period of about 10 days.

In our solar system, the accretion and metallic core segregation in the terrestrial planets occurred about 10 Myr after the formation of the solar system (Yin et al., 2002; Kleine et al., 2002; Jacobsen 2005). The period of time before 10 Myr is thought to include accretion of planetesimals and planetary embryos, and that oligarchic accretion then was largely complete (~70%) by 10 Myr (Jacobsen . Initial differentiation of metal from silicates likely occurred in planetesimals or planetary embryos early in solar system formation (e.g. Ringwood 1979), as is thought to be the case with the asteroid Vesta, which appears to have a differentiated metallic core and silicate mantle.

Several processes have been suggested for core formation during accretion of planetary embryos to form the terrestrial planets. First, if cores of planetary embryos are not vaporized or severely fragmented during accretion, their large dense metal cores will sink rapidly into the growing planet. If, however, more recent dynamical models are correct, then severe fragmentation and melting of accreting planetesimals is almost unavoidable during late-stage impacts (e.g., Stevenson, 1981; Canup, 2004). Substantial efforts have been made to understand the chemistry of core formation in a highly reduced silicate magma ocean (e.g., Allègre et al. 1995, Li & Agee 1996, Righter & Drake 1999, Wade & Wood, 2005), with the result that core segregation is now widely thought to occur through metal ponding at the bottom of planetary magma oceans, and subsequent sinking of metal diapers into the deeper planetary interior.

Righter & Drake (1999) have shown experimentally that as much as 6 wt% water can be added to the silicate magma ocean before oxygen activity begins to oxidize iron metal. Therefore both high water (or other oxidizing agent) contents and relatively long residence times in the magma ocean are required to oxidize iron and prevent its sinking into a core.

Planetesimals and planetary embryos are therefore more likely than larger bodies to oxidize iron metal during differentiation. The small gravity on planetesimals contributes to slow sinking velocities for the metal droplets, and therefore long times available for oxidation in the silicate portion of the body. Metal droplets smaller than about 2 cm will take days to settle out of a magma ocean on a small body. Simple calculations (Elkins-Tanton and Seager, submitted) indicate that if metal within planetesimals or planetary embryos is broken into droplets that are centimetric or smaller in size, they are likely to be fully oxidized as long as oxidizing agents are available.

The end member of iron oxidation in planetary embryos is a new kind of terrestrial planet: one without an iron core. This planet would be formed only from planetary embryos without metallic iron, or it would oxidize iron in its own formation because of a highly oxidizing bulk composition. It would consist entirely of silicates constructed from oxides, similar to the Earth's mantle. In other words, all iron is oxidized and incorporated in the silicates. We explore this idea further in another paper (Elkins-Tanton & Seager submitted).

**Conclusions**

We have considered the possible mass and composition range of super Earth exoplanet atmospheres (for planets ranging in mass from 1 to 30 Earth masses) based on outgassing. The outgassed atmospheres are based on the range of primitive and differentiated meteorites as building blocks for rocky planets. Reducing conditions during accretion can produce a planet with a metallic core and water-rich atmosphere, while oxidizing conditions can produce planets without a metallic core but with a hydrogen-rich atmosphere. Planets with mantles that solidified from magma oceans will retain a fraction of water in their silicate mantles sufficient to speed mantle convection and later volcanism, but small in comparison to the water and carbon compounds degassed into the young atmosphere.

Our outgassing planet models predict two kinds of planets not yet identified in our solar system: A planet consisting of silicate rock with no metallic core (Elkins-Tanton & Seager, submitted to ApJ), and a planet with an exclusively deep water surface from its initial solidification.

Water and carbon compounds will dominate atmospheres built via the models discussed here. Hydrogen in the atmosphere will not amount to more than ~6 mass% of the planet unless it is obtained through a process other than degassing, and helium will be found only in traces. Nitrogen, though in very low concentration in these planetary building materials, is sufficient to build the nitrogen-based atmosphere of the Earth. Primordial helium is present only in trace amounts even in planets with high enough surface gravities to retain this light element. . If there is a way to observe He in hydrogen-rich exoplanet atmospheres, one could distinguish between atmospheres created from outgassing alone and atmospheres formed from accretion of nebular gases.

We have shown that based on the wide range of meteorite compositions the mass and range of compositions of super-Earth atmospheres from outgassing will be richly diverse. The initial atmospheric mass could range from 1 percent of the planet's total mass or less, to a few percent, and even up to over 20% in two extreme cases. The planet's initial composition includes may be dominated by carbon compounds, hydrogen, or water, depending on the amount of accreted water-rich particles or planetesimals and the presence or absence of a magma ocean. Until observations of super Earth atmospheres are possible, this wide range of possibilities together with photochemical and atmospheric escape models must be considered for a starting point for super Earth atmospheric properties.

**Acknowledgements**

We thank Dimitar Sasselov for useful discussions and the referee for constructive suggestions. This work is supported by award 0747154 from the NSF Division of Astronomical Sciences

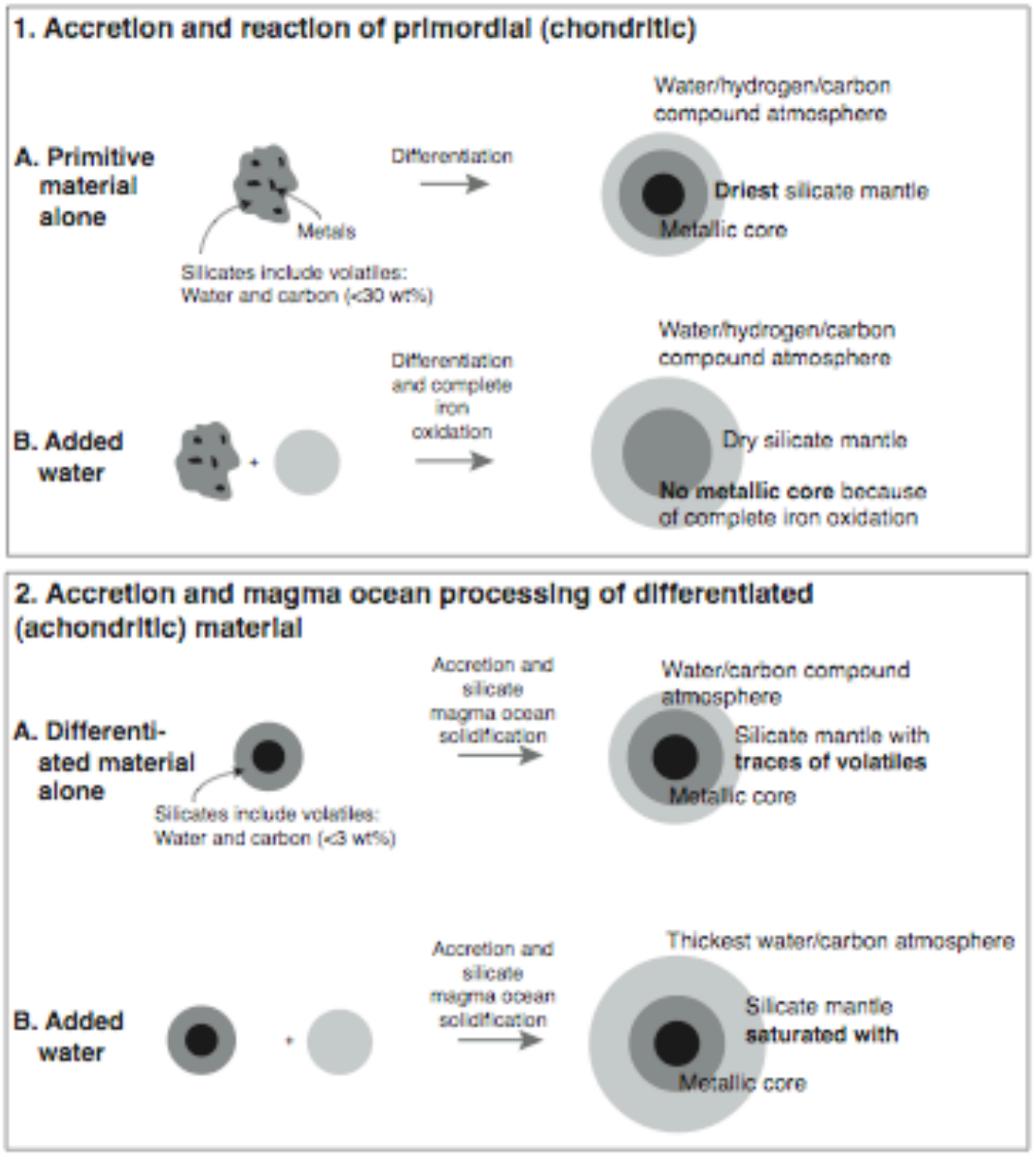

**Figure 1**. Schematic of the four atmosphere-outgassing models considered here.

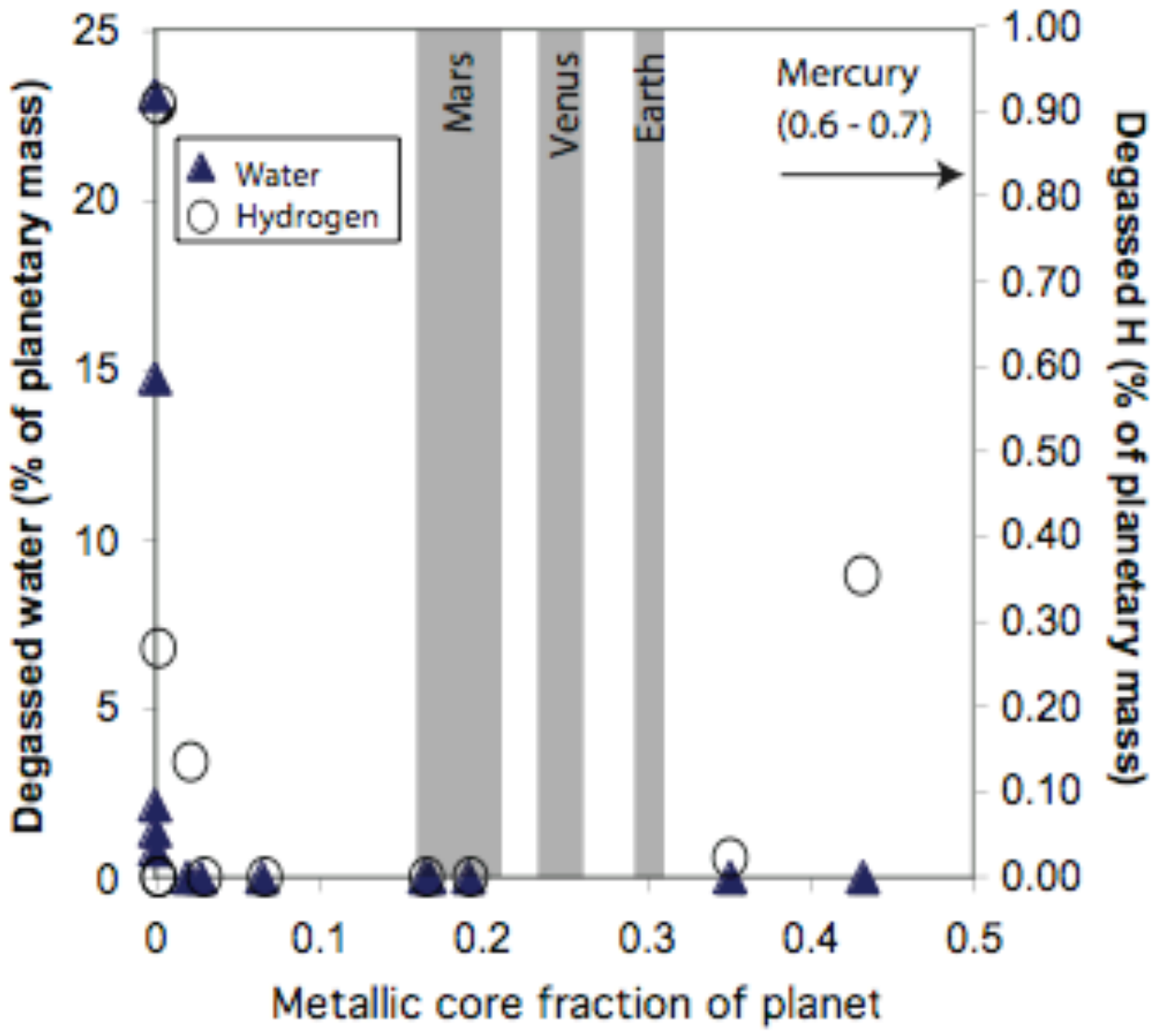

**Figure 2**. The mass fraction of water and hydrogen atmospheres vs. the mass fraction of a metallic core for planets built from chondritic (primitive) material. Water (left axis, triangles) and hydrogen (right axis, circles) released from primitive material after iron is oxidized by all existing water, as a function of remaining metallic iron fraction in the planet. Likely mass fractions of metallic iron cores of terrestrial planets are shown for comparison. The mass fraction of water and hydrogen atmospheres are upper limits, assuming that all existing water reacts with iron in the planet.

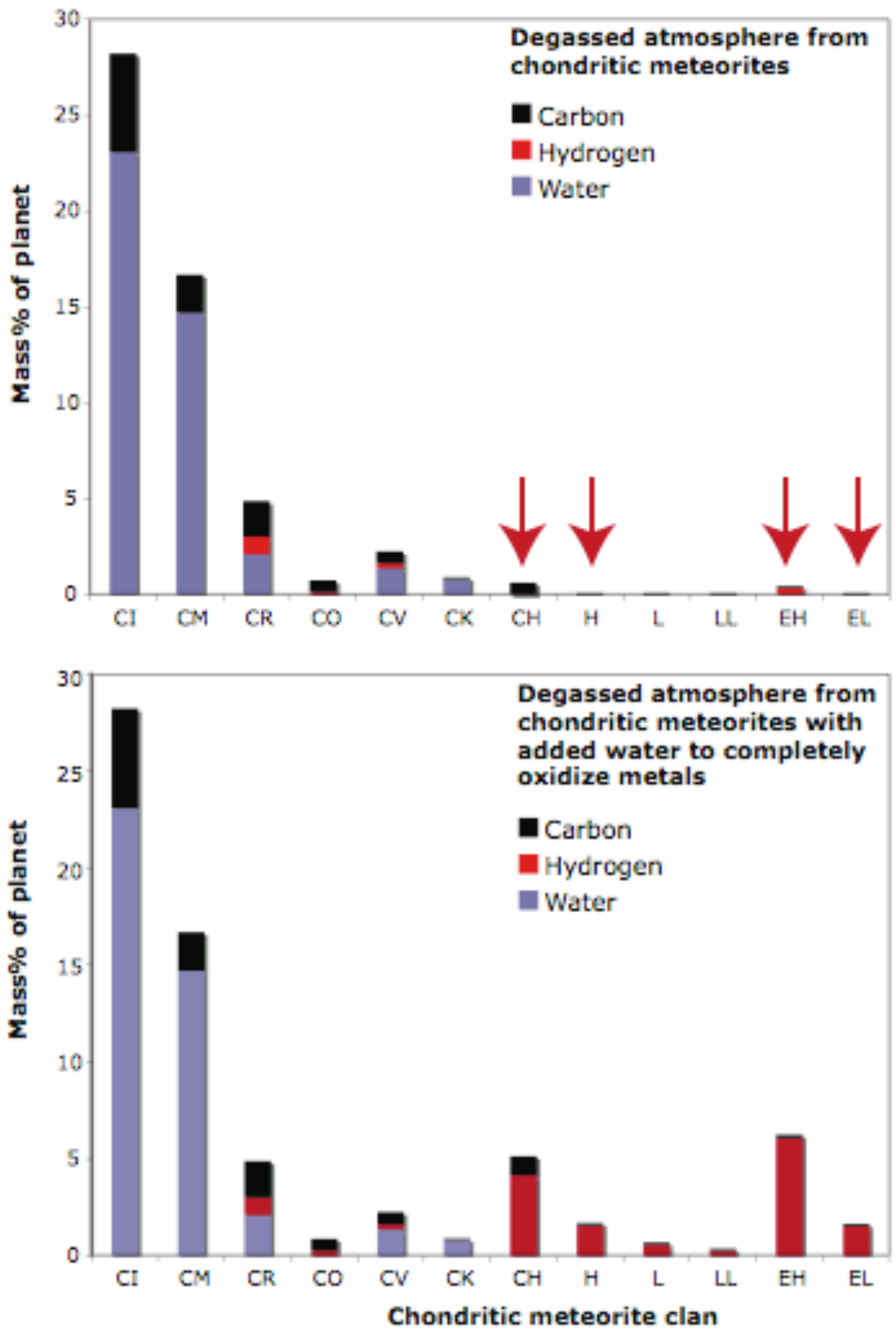

**Figure 3**. **Top**: The maximum mass of atmospheric constituents that would be degassed from a planet built of the bulk compositions of each of the common clans of chondritic meteorites. These models assume that any water present would react with any metallic iron to produce iron oxide+hydrogen. Red arrows mark compositions with sufficient metallic iron to create the cores of Mars, Venus, or the Earth.

**Bottom**: The same calculations, but with added water to allow all metallic iron to react into iron oxide: hydrogen, creating the maximum hydrogen atmosphere but no metallic core.

All compositions from Brearley & Jones (1998) and Hutchison (2004).

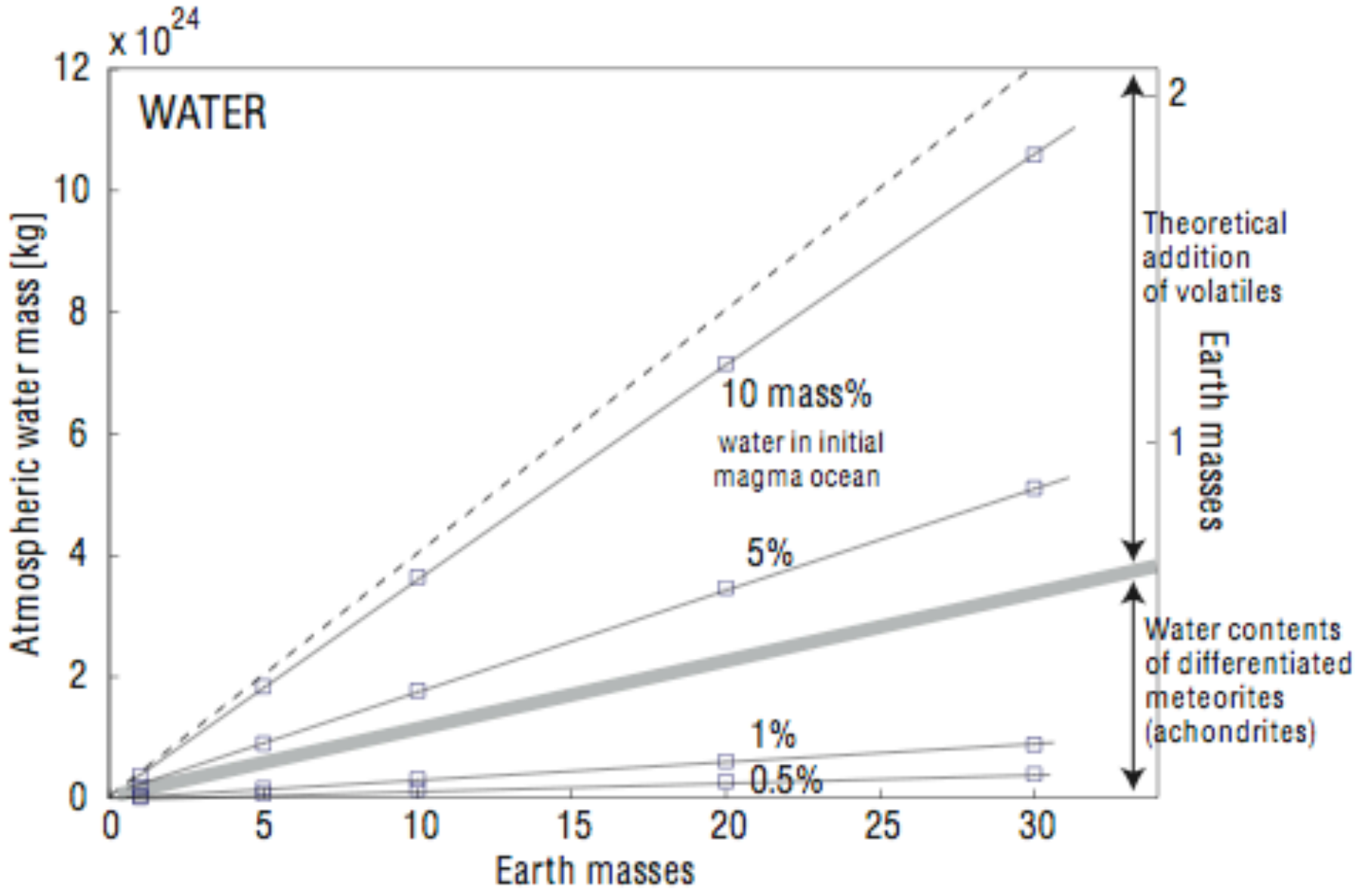

**Figure 4**. Atmospheric water mass as a function of planet mass for differentiated (achondritic) mantle compositions processed through a magma ocean. The grey line marks 3 mass% initial water content, the maximum water content measured in achondritic meteorites. The relations are not linear: water is held in solidifying minerals according to the water content of the silicate liquid from which they crystallize; higher atmospheric pressure allows higher water content in silicate liquids; atmospheric pressure grows with planetary mass; therefore more massive planets hold more water in silicate liquids, and partition more water into mantle minerals. For comparison, the dotted line indicates a linear relationship between a 10% initial water content and final atmospheric content.

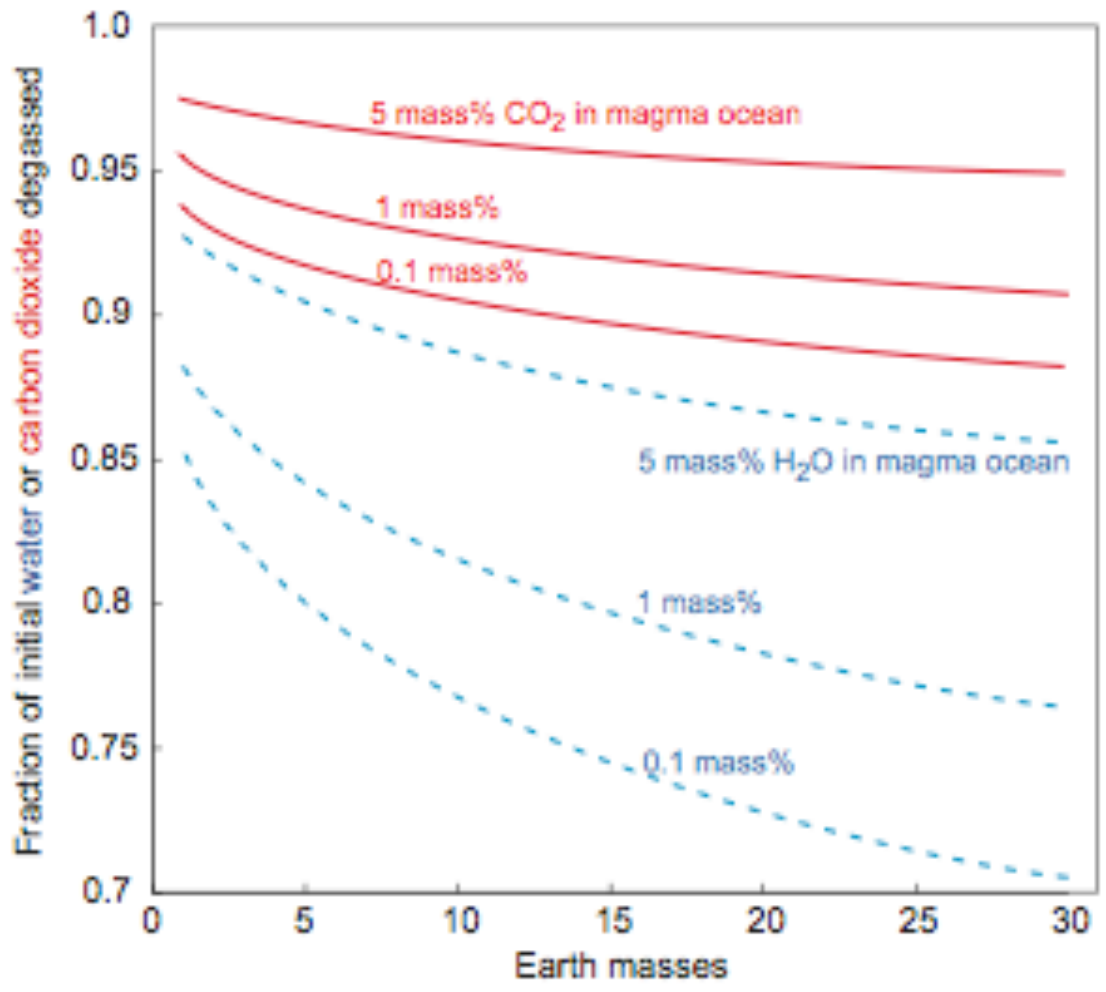

**Figure 5.** Fraction of initial volatiles in the magma ocean that are degassed into the initial atmosphere, as a function of mass % of initial volatiles and of planetary mass. The different lines are for different initial bulk magma ocean compositions, as labeled. Carbon is retained in silicate mantle minerals in much lower concentrations than is water, and so higher fractions degas into the atmosphere. Larger planets with higher atmospheric pressures retain more volatiles in their interiors. If a planet begins with equal mass fractions of carbon dioxide and water, the carbon dioxide will degas earlier and more completely than the water, producing a carbon-dominated atmosphere.

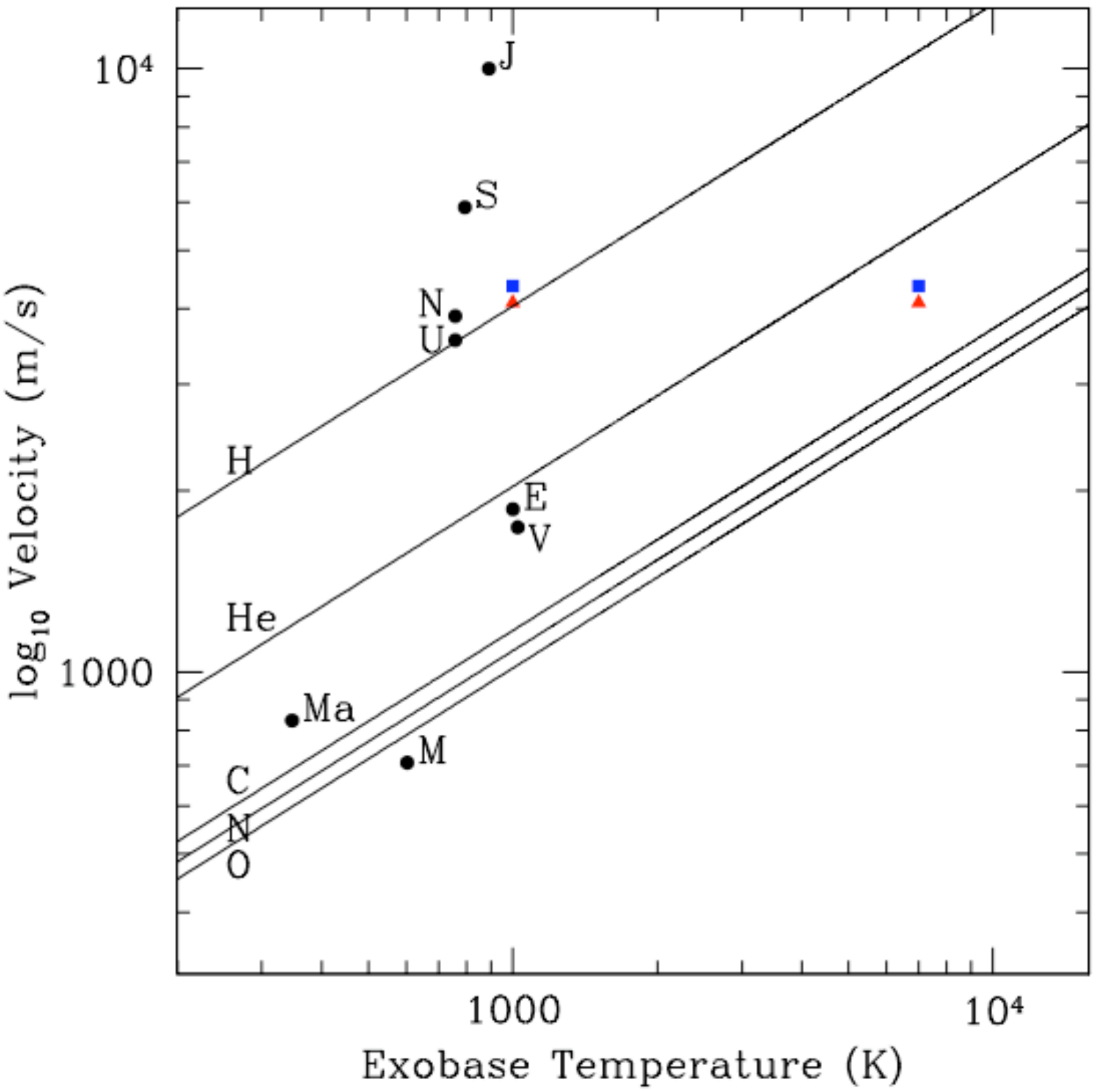

**Figure 6.** Planet escape velocity compared to thermal velocity of different elements, as a function of the exobase temperature. The solid curves are the thermal velocity of different elements and the symbols represent the escape velocity of a planet. With respect to thermal escape, planets will retain all gases with thermal velocities below the planet escape velocities. Note that the escape velocity is divided by 6 in order to represent escape over the lifetime of the planet. The solar system planets are noted with their initials.The triangle represents a 10 Earth-mass planet with a similar differentiated composition as Earth and the squares represent a 10 Earth mass planet composed of 25% water by mass outer layer, a 52.5% by mass silicate mantle and a 22.5% by mass iron core (Seager et al. 2007). The 1000 K and 10,000 K exobase temperatures are meant to represent that of super Earths in short-period orbits around M stars and Sun-like stars respectively. Super Earths in longer period orbits are expected to have lower exobase temperatures. While nonthermal atmospheric escape processes may be significant, this figure illustrates that massive super Earths can retain hydrogen against thermal escape.

**Table 1**. Models and generalized results

| | No additional volatiles<br>Process | Resulting Bulk Initial Atmosphere | Additional volatiles<br>Process | Resulting Bulk Initial Atmosphere |
|---|---|---|---|---|
| Primitive starting material (chondritic) | Metallic iron reacts with water to form hydrogen and iron oxide until either metallic iron or water is exhausted | Sufficient metallic iron remaining to form a planetary core: Atmosphere is H and carbon compounds only, with trace He and N<br>No metallic iron remaining: Water + carbon compounds ± H, with trace He and N | Water is added from an external source until all metallic iron is oxidized | Water, H, or mix plus carbon compounds, with trace He and N |
| Previously differentiated starting material (achondritic) | Molten achondritic silicate planetary mantle solidifies, partitioning volatiles between silicate minerals, evolving liquids, and atmosphere | Water and carbon compounds, with trace He and N | Additional water is added to achondritic composition and magma ocean solidification proceeds | Water and carbon compounds, with trace He and N |

**Table 2**. Chondrite classes: Degassed water, hydrogen, and carbon as a percentage of planetary mass (carbon will be degassed as carbon dioxide or other species, as discussed in the text).

| | Model 1A: Atmosphere resulting from original composition | | | | Model 1B: Atm. formed after water is added to oxidize all metals | | | |
|---|---|---|---|---|---|---|---|---|
| | as a perc. of solid planet | | | | as a perc. of solid planet | | | |
| Chondrite class | water (mass%) | H (mass%) | N (mass%) | C (mass%) | water (mass%) | H (mass%) | N (mass%) | C (mass%) |
| CI | 23.08 | 0.00 | 0.19 | 5.13 | 23.08 | 0.00 | 0.19 | 5.13 |
| CM | 14.70 | 0.00 | 0.18 | 1.98 | 14.70 | 0.00 | 0.18 | 1.98 |
| CR | 2.13 | 0.91 | 0.00 | 1.85 | 2.13 | 0.91 | 0.00 | 1.85 |
| CO | 0.00 | 0.14 | 0.01 | 0.61 | 0.00 | 0.28 | 0.01 | 0.62 |
| CV | 1.38 | 0.27 | 0.01 | 0.62 | 1.38 | 0.27 | 0.01 | 0.62 |
| CK | 0.81 | 0.00 | 0.00 | 0.10 | 0.81 | 0.00 | 0.00 | 0.10 |
| CH | 0.00 | 0.02 | 0.00 | 0.60 | 0.00 | 4.19 | 0.00 | 0.96 |
| H | 0.00 | 0.00 | 0.00 | 0.10 | 0.00 | 1.58 | 0.00 | 0.12 |
| L | 0.00 | 0.00 | 0.00 | 0.10 | 0.00 | 0.59 | 0.00 | 0.11 |
| LL | 0.00 | 0.00 | 0.01 | 0.10 | 0.00 | 0.28 | 0.01 | 0.10 |
| EH | 0.00 | 0.35 | 0.00 | 0.10 | 0.00 | 6.09 | 0.00 | 0.18 |
| EL | 0.00 | 0.00 | 0.00 | 0.10 | 0.00 | 1.53 | 0.00 | 0.12 |